# CYBERCRIME AND COMPUTER FORENSICS IN EPOCH OF ARTIFICIAL INTELLIGENCE IN INDIA

Dr. Shikha Dhiman[*]
Sahibpreet Singh[**]

**ABSTRACT**

The advent of Artificial Intelligence (AI) has ushered in a transformative era across diverse sectors, with profound implications for the domains of criminal justice, cybersecurity, and digital forensics. AI's capacity to augment capabilities in countering cybercrimes while simultaneously introducing challenges and ethical quandaries necessitates a meticulous examination. Integration of AI within the realms of cybercrime and computer forensics mandates a judicious and balanced approach, grounded in ethical principles, established standards, and legal regulations that accord priority to human rights, privacy, and security. These regulations must underscore transparency, accountability, and fairness in the deployment of AI systems. Effective handling of AI-driven cyber threats necessitates collaboration and coordination among governments, private sector entities, civil society, academia, and technical experts. Such collaboration enables the sharing of best practices and knowledge, thus facilitating a more robust collective response to the ever-evolving landscape of cybercrimes. Education and public awareness form integral components in preparing society for an AI-driven future. Equipping law enforcement agencies, legal professionals, and forensic experts with training and resources empowers them to navigate cases involving AI technologies efficiently. Bolstering the capabilities of the criminal justice system is of paramount importance. This involves the development of pertinent legal frameworks, technical tools, forensic methodologies, evidentiary standards, and judicial procedures to accommodate the evolving spectrum of AI-enabled crimes. Innovation and research play a pivotal role in countering threats posed by AI, with a particular emphasis on the development of trustworthy AI systems that are resilient, secure, and human-centric. Such AI systems are instrumental in mitigating the malevolent applications of AI, ensuring the privacy and security of individuals and organizations alike. As AI continues to evolve and shape the digital landscape, a proactive and comprehensive approach is imperative. Embracing the opportunities that AI offers while concurrently addressing its challenges through responsible governance and ethical considerations creates a safer and more secure digital environment. This approach allows society to harness the full potential of AI while safeguarding privacy and security in the era of cybercrime and computer forensics. This legal and ethical analysis underscores the multifaceted implications of AI on privacy, security, cybercrime, and computer forensics, and offers recommendations for a judicious and ethical approach in the age of digital transformation.

---

[*] Assistant Professor, Department of Laws, Guru Nanak Dev University, Amritsar, Punjab.
[**] LLM (2023-24), Department of Laws, Guru Nanak Dev University, Amritsar, Punjab.

**INTRODUCTION**

AI is a term that encompasses various technologies that enable machines to perform tasks that typically require human intelligence, such as learning, reasoning, decision-making, and problem-solving. Artificial intelligence (AI) has grown exponentially in recent years due to the abundance of data, large-scale computing power, and advanced algorithms. AI has been applied in various domains, such as healthcare, education, entertainment, finance, and transportation, to improve efficiency, accuracy, and convenience. However, AI also poses significant challenges and risks for privacy, security, and forensics, especially in relation to cybercrime and computer forensics.[1] Cybercrime is a term used to describe any criminal activity that takes place on, or involving, a computer, network, or digital device. Cybercrime can take various forms, such as hacking, phishing, malware, ransomware, identity theft, fraud, cyberstalking, cyberbullying, cyberterrorism, and cyberwarfare.[2] Cybercrime can cause serious harm to individuals, organizations, and society at large, such as financial losses, reputational damage, emotional distress, physical injury or even death.[3] Computer forensics is a crucial branch of digital forensics that deals with the identification, preservation, extraction, analysis, and presentation of digital evidence from computers or other digital devices.[4] It aims to support investigations of cybercrime or other incidents where digital evidence is relevant. Computer forensics involves various techniques and tools to collect and analyze data from storage media, memory, network traffic, logs, or applications.[5] It also requires adherence to legal and ethical principles to ensure the validity and admissibility of the evidence.

AI can have both positive and negative impacts on privacy, security, and forensics in the age of cybercrime and computer forensics. On one hand, AI can enhance privacy protection by enabling encryption[6], anonymization[7], or differential privacy[8] techniques that can prevent unauthorized access or disclosure of sensitive data. AI can also improve security by enabling detection, prevention, or mitigation techniques that can counteract cyberattacks or reduce their damage. AI can also facilitate forensics by enabling automation, analysis, or visualization techniques that can speed up the process of evidence collection and examination.

On the other hand, AI can also threaten privacy by enabling surveillance, profiling, or inference techniques that can monitor or predict the behavior or preferences of individuals or groups without their consent or knowledge. AI can also compromise security by enabling attacks, evasion, or manipulation techniques that can exploit vulnerabilities or deceive defenses in systems or networks. AI can also challenge forensics by enabling obfuscation, anti-forensics, or deepfakes techniques that can hide or alter the traces or sources of cybercrime or generate synthetic media content that can mislead or impersonate. According to a report by McAfee (2020), the global cost of cybercrime was estimated to be $945 billion in 2020[9], which is more than 1% of the global GDP[10].

Therefore, it is important to understand the impact of AI on privacy and security in the age of cybercrime and computer forensics. This research article aims to provide a comprehensive overview of the current state-of-the-art and future directions of AI in relation to privacy protection, security enhancement, and forensic investigation.

**SCOPE OF AI IN THE EPOCH OF CYBERCRIME AND COMPUTER FORENSICS**

Artificial Intelligence (AI) has deeply revolutionised various fields, including criminology, social engineering, psychology, computer science and robotics. As artificial intelligence systems adopt the security vulnerabilities of conventional computer systems, the worry about new forms of cyberattacks amplified by AI is on the rise too. AI is deeply connected to physical space (e.g., self driving vehicles, intelligent virtual assistants), so AI-related crime can harm people physically, beyond the cyberspace. AI crimes can be broadly classified into following categories:

i. AI as a tool crime and
ii. AI as a target crime.[11]

Practically, AI has brought a lot of benefits and advantages to humanity. Globally, governments are considering the active use of AI systems and applications to help them achieve their activities and more concretely to facilitate the identification and prediction of crime. National security and intelligence agencies have also realized the potential of AI technologies to support and achieve national and public security objectives. While AI has potential to enhance the capabilities in countering cybercrime and improving computer forensics, it also presents new challenges in terms of privacy and security[12]. It's crucial for ongoing policy initiatives of international organizations to consider certain aspects in the law-making process in the field of cybercrime.[13]

The development of AI has had a significant impact on both cybercrime and computer forensics. On one hand, AI has enabled new forms and modes of cybercrime, such as identity theft, phishing, spamming, botnets, ransomware, cyberattacks, and cyberterrorism. AI has also enhanced the capabilities and sophistication of cybercriminals, who can use AI techniques to evade detection, encrypt data, generate malware, impersonate victims, and manipulate information.[14] On the other hand, AI has also facilitated new methods and tools for computer forensics, such as data mining, pattern recognition, natural language processing, image processing, machine learning, and neural networks. AI has also improved the efficiency and accuracy of computer forensic analysts, who can use AI techniques to automate tasks, extract features, classify data, identify anomalies, and generate reports.[15] The reign of AI, cybercrime, and computer forensics is still unfolding. As technology evolves, so do the opportunities and challenges for both criminals and investigators.

**IMPACT OF AI ON PRIVACY**

Artificial intelligence (AI) is a rapidly evolving technology that has the potential to transform various aspects of human life, such as health, education, entertainment, and commerce. However, AI also poses significant challenges to the protection of privacy, which is a fundamental human right and a core value of democratic societies. Privacy can be defined as the right of individuals to control their personal information and to decide how, when, and by whom it is collected, processed, and shared.[16] Artificial intelligence (AI) systems are often based on massive amounts of personal information, which raises questions about data collection, data processing, and data storage. Moreover, AI systems can use personal data in ways that can intrude on privacy interests by enabling new and unanticipated inferences and predictions, such as facial recognition, behavioral analysis, and profiling.[17] The impact of AI on privacy can be analyzed from different perspectives, viz legal and ethical.

   A.  *Legal Perspective*

India has recently enacted a comprehensive privacy law that governs the processing of digital personal data of Indian residents by both public and private entities. The Digital Personal Data Protection Act, 2023 (DPDP Act) is a landmark legislation that was passed by the Parliament of India on August 31, 2023.[18] The DPDP Act, 2023 is based on the principles of consented, lawful and transparent use of personal data, purpose limitation, data minimisation, data accuracy, storage limitation, reasonable security safeguards, and accountability. The DPDP Act also introduces some innovative features and concepts, such as data fiduciaries, data principals, data trusts, consent managers, sandboxing, and social media intermediaries. The DPDP Act establishes a Data Protection Board (DPB) that has the power to enforce the law and issue codes of practice and guidance.[19]

Some key terms used in the Digital Personal Data Protection Act, 2023 are discussed as under:

i. **Digital personal data**[20]: Personal data in digital form. The Act applies to the processing of digital personal data within and outside the territory of India, subject to certain conditions and exemptions[21].

ii. **Data principal**[22]: The individual to whom the digital personal data relates. The Act introduces duties for data principals and imposes a penalty up to INR 10,000 for any breach of duty.

iii. **Data fiduciary**[23]: Any person, including the State, a company, any juristic entity or any individual who alone or in conjunction with others determines the purpose and means of processing of digital personal data. The Act imposes various obligations on data fiduciaries, such as obtaining consent, providing notice, ensuring security safeguards, conducting data privacy impact assessments, etc.

iv. **Data processor**[24]: Any person, including the State, a company, any juristic entity or any individual who processes digital personal data on behalf of a data fiduciary. The Act requires data fiduciaries to engage data processors through a contract and hold them accountable for any non-compliance.

v. **Consent**[25]: A clear, specific, informed and free expression of will by the data principal to allow the processing of his or her digital personal data. The Act also provides for consent withdrawal, verifiable parental consent, and consent managers.

vi. **Notice**[26]: A concise, transparent, intelligible and easily accessible communication from the data fiduciary to the data principal about the purpose, means and manner of processing of digital personal data. The Act specifies the contents and form of notice and requires it to be provided at the time of collection or as soon as possible thereafter.

vii. **Grounds of processing**[27]: The lawful bases for processing digital personal data under the Act. These include consent, contract, legal obligation, vital interest, public function and reasonable purpose.

viii. **Data protection officer**[28]: An individual appointed by the data fiduciary to perform functions such as advising on compliance, monitoring adherence to the Act and codes of practice, liaising with the Data Protection Board (DPB), etc. The Act mandates certain categories of data fiduciaries to appoint a data protection officer.

ix. **Data Protection Board**[29]: A statutory body established under the Act to perform functions such as enforcing the law, issuing codes of practice and guidance, conducting inquiries and investigations, imposing penalties and compensation, etc. The DPB consists of a chairperson and six whole-time members appointed by the central government.

x. **Data trusts**: A mechanism for collective governance of digital personal data by a group of stakeholders who have a common interest in its use and benefit. The Act empowers the DPB to recognise and certify data trusts as per prescribed criteria.[30]

The Digital Personal Data Protection Act, 2023 (DPDPA, 2023) differs from other existing or proposed laws on privacy protection in India in several ways:

1. The IT Act, 2000 covers not only electronic records, but also any information in any material form that is accessed or processed by a computer, computer system or computer network.[31] The DPDPA, 2023 covers all forms of personal data, whether in electronic or non-electronic form, that can identify a natural person or is linked or linkable to a natural person.[32]

2. The Aadhaar Act, 2016 regulates the collection and use of Aadhaar number and biometric information for the purpose of establishing identity of an individual for any purpose.[33] The DPDPA, 2023 regulates the collection and use of all types of personal data for any lawful purpose, subject to the consent of the data principal (the person to whom the data relates) and the compliance with the principles and obligations laid down in the Act.[34]

3. The Right to Information Act, 2005 grants citizens the right to access information held by public authorities, subject to certain exemptions and restrictions.[35] The DPDPA, 2023 grants individuals the right to access their own personal data held by any data fiduciary (the person, company or government entity who processes data), subject to certain conditions and exceptions. The DPDPA, 2023 also grants individuals the right to correction, erasure, portability and restriction of processing of their personal data.[36]

AI poses both challenges and opportunities for privacy protection in India:

i. AI can enable better delivery of public services, enhance security and safety, improve health care and education outcomes, etc., by using personal data or non-personal data to provide personalized and efficient solutions.[37]

ii. AI can pose risks such as discrimination, bias, manipulation, surveillance, etc., by using personal data or non-personal data to influence or harm individuals or groups.

iii. AI can also raise ethical issues such as transparency, accountability, human dignity, autonomy, etc., by using personal data or non-personal data to make decisions or actions that affect individuals or society.

However, these challenges can be addressed in different ways, which are mentioned as under:

A. Adopting ethical principles and frameworks for AI development and deployment that respect human rights and values.[38]

B. Ensuring human oversight and intervention in AI systems that involve personal data or non-personal data processing or decision making.[39]

C. Enhancing transparency and explainability of AI systems that use personal data or non-personal data to provide information or services.[40]

D. Strengthening accountability and liability of AI systems and their developers or operators for any privacy violations or harms caused by using personal data or non-personal data.

### B. *Ethical Perspective*

From an ethical perspective, the impact of AI on privacy depends largely on the moral principles and values that guide the design, development, and use of AI systems. Ethics can be defined as the study of what is right or wrong, good or bad, in human conduct.[41] Ethics can help to evaluate the social and human implications of AI systems beyond legal compliance or technical feasibility. Ethics can also help to identify and address potential conflicts or trade-offs between different values or interests that may arise from AI systems. For example,

i. How to balance privacy with security, efficiency, or innovation?
ii. How to respect individual autonomy with social responsibility or collective welfare?
iii. How to ensure fairness with accuracy, transparency, or accountability?

Several ethical frameworks and principles have been proposed to guide the development and use of AI systems in a responsible and trustworthy manner. For example,

a) The IEEE Global Initiative on Ethics of Autonomous and Intelligent Systems has developed Ethically Aligned Design: A Vision for Prioritizing Human Well-being with Autonomous and Intelligent Systems , which outlines eight general principles for ethical AI: human rights, well-being, data agency, effectiveness, transparency, accountability, awareness of misuse, and competence.[42]

b) The High-Level Expert Group on Artificial Intelligence appointed by the European Commission has published Ethics Guidelines for Trustworthy AI, which defines seven key requirements for ethical AI: human agency and oversight, technical robustness and safety, privacy and data governance, transparency, diversity, non-discrimination and fairness, societal and environmental well-being, and accountability.[43]

c) The Partnership on AI is a multi-stakeholder initiative that aims to promote best practices and standards for ethical AI. It has adopted Tenets that reflect its shared vision and goals: safety-critical AI; fairness; transparency; collaboration; accountability; value alignment; human control; social benefit; non-subversion; well-being.[44]

The ethical perspective on the impact of AI on privacy highlights that privacy is not only a legal right but also a moral value that deserves respect and protection. Privacy is essential for human dignity, autonomy, identity, expression, and relationships. Privacy also enables individuals to exercise other rights and freedoms, such as freedom of thought, opinion, expression, association, and participation.[45] Therefore, AI systems should be designed and used in ways that uphold and enhance privacy as a core value and not undermine or erode it.

**CYBERCRIME VIS-A-VIS SECURITY**

Cybercrime is a term that encompasses various forms of illegal and harmful activities that involve the use of computers, networks, or digital devices. Cybercrime can affect individuals, organizations, and nations by compromising their privacy, security, and integrity of data and systems.[46] Some examples of cybercrime are hacking, phishing, identity theft, cyber fraud, cyber terrorism, cyber espionage, cyber bullying, and cyber stalking.[47]

Security is a term that refers to the protection of information, systems, and assets from unauthorized access, use, modification, or destruction. Security can be achieved by implementing various measures such as encryption, authentication, authorization, firewalls, antivirus software, backup systems, and security policies. Security can also be enhanced by raising awareness and educating users about the risks and best practices of using cyberspace. The relationship between security and cybercrime is complex and dynamic.[48] On one hand, security aims to prevent and mitigate cybercrime by providing safeguards and countermeasures against potential threats and attacks. On the other hand, cybercrime challenges and undermines security by exploiting vulnerabilities and loopholes in existing systems and technologies. Moreover, cybercrime can also leverage artificial intelligence (AI) to enhance its capabilities and sophistication.[49]

AI is a term that refers to the simulation of human intelligence processes by machines, especially computer systems. AI can perform tasks such as learning, reasoning, problem-solving, decision-making, natural language processing, computer vision, speech recognition, and machine learning.[50] AI can also create new forms of content such as images, videos, texts, music, and code.

AI has a significant impact on both security and cybercrime. AI has the potential to be utilized for both beneficial and detrimental purposes in the digital realm. On the positive side, AI can improve security by enabling faster and more accurate detection and response to cyberattacks[51], generating alerts and recommendations for users[52], identifying new strands of malware[53], and protecting sensitive data[54]. On the negative side, AI can also facilitate cybercrime by creating more complex, adaptable, and malicious software[55], generating fake or misleading content such as deepfakes[56], impersonating or manipulating users[57], and bypassing or compromising security systems[58]. Therefore, it is essential to understand the implications and challenges of AI for security and cybercrime in the age of digital transformation.

**AI AND COMPUTER FORENSICS**

Computer forensics is the process of collecting, preserving, analyzing, and presenting digital evidence from various sources, such as computers, mobile devices, networks, cloud services, and the Internet of Things (IoT). Computer forensics plays a vital role in investigating cybercrimes, such as hacking, phishing, fraud, identity theft, cyberterrorism, and cyberwarfare. However, computer forensics faces many challenges in the age of big data, such as the increasing volume, variety, velocity, and veracity of digital data; the complexity and diversity of digital devices and platforms; the encryption and obfuscation techniques used by cybercriminals; the legal and ethical issues related to privacy and security; and the lack of standardization and validation of forensic tools and methods[59].

One of the main applications of AI in computer forensics is to use machine learning techniques to extract relevant features and patterns from large and heterogeneous datasets of digital evidence. Machine learning can also help to classify, cluster, and correlate different types of digital evidence, such as files, emails, images, videos, logs, and network traffic. Moreover, machine learning can assist in identifying anomalies, outliers, and suspicious activities in digital data, as well as in generating hypotheses and explanations for forensic investigations. Furthermore, machine learning can enhance the accuracy and reliability of forensic tools and methods by providing feedback and evaluation mechanisms.

Artificial intelligence (AI) is a branch of computer science that aims to create machines and systems that can perform tasks that normally require human intelligence, such as learning, reasoning, decision making, problem solving, and natural language processing . AI has been applied to various domains and industries, such as healthcare, education, finance, entertainment, and security . AI can also assist computer forensics in overcoming some of the challenges mentioned above by providing automated, efficient, accurate, and reliable solutions for digital evidence mining and analysis.[60]

AI can help computer forensics in several ways:

i. AI can help in data acquisition and preservation by using techniques such as data compression, deduplication, hashing, encryption, and authentication to reduce the size of data, eliminate redundant data, verify the integrity of data, protect the confidentiality of data, and ensure the admissibility of data in court.[61]

ii. AI can help in data extraction and preprocessing by using techniques such as optical character recognition (OCR), speech recognition, natural language processing (NLP), image processing, video processing, audio processing, and signal processing to convert unstructured data into structured data; to extract relevant information from text, speech, images, videos, audio files, and signals; to filter out noise and irrelevant data; to segment data into meaningful units; to normalize data into standard formats; and to enrich data with metadata and annotations.[62]

iii. AI can help in data analysis and interpretation by using techniques such as machine learning (ML), deep learning (DL), pattern recognition, classification, clustering, anomaly detection, association rule mining, sentiment analysis, topic modeling, text summarization, face recognition, fingerprint recognition, iris recognition, voice recognition, handwriting recognition, behavior analysis, emotion analysis, and network analysis to discover hidden patterns, relationships, trends, anomalies, associations, sentiments, topics, summaries, identities, behaviors, emotions, and networks from data; to classify data into predefined categories; to cluster data into similar groups; to detect outliers and anomalies in data; to infer rules and associations from data; to generate insights and hypotheses from data; to support decision making and problem solving from data; and to present results in a clear and understandable way.[63]

iv. AI can help in data evaluation and validation by using techniques such as explainable AI (XAI), interpretable AI (IAI), transparent AI (TAI), trustworthy AI (TAI), ethical AI (EAI), adversarial learning (AL), evaluation metrics (EM), benchmark datasets (BD), cross-validation (CV), testing frameworks (TF), peer review (PR), reproducibility (RP), repeatability (RT), robustness (RB), reliability (RL), accuracy (AC), precision (PR), recall (RC), F1-score (F1), confusion matrix (CM), receiver operating characteristic curve (ROC), area under the curve (AUC), mean absolute error (MAE), mean squared error (MSE), root mean squared error (RMSE), r-squared (R2), and p-value (PV) to explain how AI models work and why they produce certain outputs; to make AI models more understandable and transparent to humans; to ensure that AI models are trustworthy and ethical and do not harm or discriminate against humans; to defend AI models against adversarial attacks and manipulation; to measure the performance and quality of AI models using various metrics and datasets; to validate and verify the results of AI models using various methods and frameworks; to review and critique the methods and results of AI models by experts and peers; to ensure that AI models are reproducible and repeatable by other researchers and practitioners; to test the robustness and reliability of AI models under different conditions and scenarios; to assess the accuracy and precision of AI models in predicting the correct outputs; to measure the recall and F1-score of AI models in retrieving the relevant outputs; to visualize the confusion matrix and ROC curve of AI models to show the trade-off between true positives, false positives, false negatives, and true negatives; to calculate the AUC of AI models to show the overall performance of binary classifiers; to compute the MAE, MSE, RMSE, R2, and PV of AI models to show the error, variance, and significance of regression models.[64]

AI can also benefit from computer forensics by using the data and knowledge generated by computer forensics as inputs and feedbacks for AI models to improve their learning and adaptation capabilities; by using the methods and tools developed by computer forensics as references and benchmarks for AI models to enhance their quality and standards[65]; and by using the challenges and problems faced by computer forensics as opportunities and motivations for AI models to innovate and evolve.[66] However, AI also poses some risks and limitations for computer forensics, such as:

i. AI can be used by cybercriminals to create more sophisticated and stealthy attacks, such as malware, ransomware, botnets, phishing, social engineering, identity theft, deepfakes, fake news, and disinformation campaigns.

ii. AI can be used by cybercriminals to evade detection and attribution, such as using encryption, obfuscation, steganography, watermarking, proxy servers, virtual private networks (VPNs), Tor network, dark web, cryptocurrency, blockchain, and zero-knowledge proofs.

- iii. AI can be used by cybercriminals to tamper with or destroy digital evidence, such as using anti-forensics techniques, data wiping tools, file shredders, data corruption tools, data fabrication tools, and data poisoning attacks.
- iv. AI can be affected by human biases and errors, such as confirmation bias, overfitting, underfitting, sampling bias, selection bias, measurement bias, algorithmic bias, cognitive bias, and ethical bias.
- v. AI can be vulnerable to adversarial attacks and manipulation, such as using adversarial examples, adversarial perturbations, adversarial patches, adversarial poisoning, adversarial vasion, adversarial trojans, adversarial backdoors and adversarial reprogramming.
- vi. AI can be difficult to explain and interpret, especially for complex and nonlinear models such as neural networks, support vector machines, random forests, gradient boosting machines, and genetic algorithms.
- vii. AI can be challenging to evaluate and validate, especially for novel and unconventional models such as quantum machine learning, neuromorphic computing, spiking neural networks, reservoir computing, and cellular automata.[67]

Therefore, computer forensics needs to adopt a balanced and cautious approach towards AI by leveraging its advantages and mitigating its disadvantages; by collaborating with other disciplines and stakeholders such as law enforcement agencies, judiciary system, academia, industry, civil society, and international organizations; by following ethical principles and legal frameworks such as fairness, accountability, transparency, privacy, security, human rights, data protection laws, cybercrime laws, evidence laws; and by developing best practices and standards for using AI in computer forensics.[68]

**SUGGESTIONS**

This research article examines the impact of AI on privacy and security, especially in the context of cybercrime and computer forensics. It analyzes the legal and ethical issues that arise from the use of AI in these areas, and the existing instruments that regulate or address them. It also highlights some of the current trends and challenges that AI poses to the criminal justice system and the society at large. The future of AI will pose new ethical, legal, and social issues for privacy and security in cyberspace. The role of computer forensics will become more crucial and complex in the age of AI.

Based on this analysis, the following recommendations are suggested for enhancing privacy and security in the age of AI:

i. Develop and implement global standards and norms for responsible and ethical use of AI in criminal justice, cybersecurity, and digital forensics. These should include principles such as transparency, accountability, fairness, human oversight, and data protection.
ii. Strengthen the cooperation and coordination among different stakeholders, such as governments, international organizations, academia, industry, civil society, and experts, to share best practices, exchange information, and provide guidance on AI-related issues.
iii. Promote awareness and education among the public and the professionals about the benefits and risks of AI, and the rights and responsibilities that come with it. This should include providing training and resources for law enforcement, prosecutors, judges, lawyers, forensic experts, and other relevant actors on how to deal with AI-related cases.
iv. Enhance the capacity and capability of the criminal justice system to prevent, detect, investigate, prosecute, and adjudicate AI-related crimes. This should include developing and adopting appropriate legal frameworks, technical tools, forensic methods, evidentiary standards, and judicial procedures.
v. Foster innovation and research on AI that can improve privacy and security, as well as counter the threats posed by malicious AI. This should include supporting the development of trustworthy AI systems that are robust, resilient, secure, and human-centric.

**CONCLUSION**

It can be articulated that the emergence of Artificial Intelligence (AI) has brought about a profound transformation in various sectors, including criminal justice, cybersecurity, and digital forensics. AI holds the potential to significantly enhance our capabilities in combating cybercrimes and improving overall security. However, it also introduces a range of challenges and ethical dilemmas that require careful consideration. The integration of AI into the realms of cybercrime and computer forensics necessitates a balanced approach. The development and use of AI should be guided by ethical principles, standards, and regulations that prioritize human rights, privacy, and security. These regulations should emphasize transparency, accountability, and fairness in the deployment of AI systems.

Collaboration and coordination among governments, private sector organizations, civil society, academia, and technical experts are critical to addressing the global nature of cyber threats facilitated by AI. Sharing best practices and knowledge will enable a more effective collective response to evolving cybercrimes. Education and awareness are essential elements in preparing society for the AI-driven future. Providing training and resources to law enforcement, legal professionals, and forensic experts will empower them to effectively handle cases involving AI technologies.

Moreover, bolstering the capabilities of the criminal justice system, including the development of appropriate legal frameworks, technical tools, forensic methods, evidentiary standards, and judicial procedures, is imperative to adapt to the changing landscape of AI-enabled crimes. Innovation and research play a vital role in countering AI-enabled threats. The development of trustworthy AI systems that prioritize security, resilience, and human-centricity is essential to mitigate the malicious uses of AI and ensure the privacy and security of individuals and organizations.

As AI continues to evolve and shape our digital environment, a proactive and comprehensive approach is necessary. By embracing the opportunities presented by AI while addressing its challenges through responsible governance and ethical considerations, a safer and more secure digital landscape can be created for all. This approach allows us to fully leverage the potential of AI while safeguarding privacy and security in the era of cybercrime and computer forensics.

# ENDNOTES

[30] PwC India. 2023. *The Digital Personal Data Protection Act, India 2023*. India: PricewaterhouseCoopers Private Limited.

[31] *Section 2(1)(t) of the IT Act,* 2000.

[32] *Chapter II of the DPDPA,* 2023.

[33] *Section 7 of the Aadhaar Act,* 2016.

[34] *Chapter II of the DPDPA,* 2023.

[35] *Chapter II of the RTI Act,* 2005.

[36] *Chapter II of the DPDPA,* 2023.

[37] Mehra, Samiksha. "The Interplay of AI, Data, and Privacy." indiaai.gov.in/article/the-interplay-of-ai-data-and-privacy.

[38] Indiaai.gov.in. "Artificial Intelligence and Privacy." indiaai.gov.in/research-reports/artificial-intelligence-and-privacy.

[39] Batra, Tuhin. "Self-Regulation in Artificial Intelligence: An Indian Perspective - Privacy Protection - India." www.mondaq.com/india/privacy-protection/1015476/self-regulation-in-artificial-intelligence-an-indian-perspective.

[40] Bhatia, Kalindhi. "The "ChatGPT Effect" – Parsing Privacy and AI Regulation in India - New Technology - India." www.mondaq.com/india/new-technology/1311384/the-chatgpt-effect--parsing-privacy-and-ai-regulation-in-india.

[41] Beauchamp, Tom L., and James F. Childress. 2019. *Principles of Biomedical Ethics*, Eighth Edition. United States: Oxford University Press.

[42] IEEE Global Initiative on Ethics of Autonomous and Intelligent Systems. 2019. *Ethically Aligned Design: A Vision for Prioritizing Human Well-being with Autonomous and Intelligent Systems*. Piscataway, NJ: IEEE.

[43] High-Level Expert Group on Artificial Intelligence. 2019. *Ethics Guidelines for Trustworthy AI*. Brussels, Belgium: European Commission.

[44] Partnership on AI. "Tenets." Partnershiponai.org.

[45] United Nations General Assembly. 1948. *Universal Declaration of Human Rights*, Resolution 217A(III). Paris, France: United Nations.

[46] Hawdon, James. 2021. "Cybercrime: Victimization, Perpetration, and Techniques." *American Journal of Criminal Justice.*

[47] Boruah, Jayanta. 2020. "Cyber Crimes and Its Legal Challenges in India." *The Journal of Legal Methodology Policy and Governance*, Vol. 2, Issue 1.

[48] Cremer, Frank, et al. 2022. "Cyber Risk and Cybersecurity: A Systematic Review of Data Availability." *The Geneva Papers on Risk and Insurance - Issues and Practice*, Vol. 47, No. 3.

[49] Lavorgna, Anita, and Thomas J. Holt. 2021. *Researching Cybercrimes: Methodologies, Ethics, and Critical Approaches*. Switzerland: Palgrave Macmillan.

[50] Sheikh, Haroon, et al. 2023. *Mission AI*. United Kingdom: Springer Nature.

[51] Buczynski, Bethany. "How AI Can Help Us Fight Cybercrime." us.norton.com/internetsecurity-emerging-threats-how-ai-can-help-us-fight-cybercrime.html.

[52] Chen, Yuxi, et al. "AI for Cybersecurity: A Research Roadmap." arxiv.org/abs/2008.01713.

[53] Eshel, Rami. "How AI Is Changing the Cybersecurity Landscape." forbes.com/sites/forbestechcouncil/2019/11/14/how-ai-is-changing-the-cybersecurity-landscape.

[54] Garg, Ankit. "How AI Can Help Protect Your Data Privacy." analyticsindiamag.com/how-ai-can-help-protect-your-data-privacy.

[55] Khandelwal, Swati. "New AI Tool Can Write Malware That Evades Antivirus Detection." thehackernews.com/2019/05/artificial-intelligence-malware.html.

[56] Klonick, Kate. "The Rise of the Deepfake and the Threat to Democracy." theguardian.com/commentisfree/2018/nov/12/deepfake-democracy-artificial-intelligence.